\documentclass[12pt]{article}

\usepackage{amsmath}
\usepackage{amssymb}
\usepackage{graphicx}

\begin{document}

\title{\bf Finite Action  Klein-Gordon Solutions  on  Lorentzian Manifolds}

\author{V.V. Kozlov and I.V. Volovich $~~~~$\\
$~~~~~~~~~$\\
Steklov Mathematical Institute\\
Russian Academy of Sciences\\
Gubkin St. 8, 119991, Moscow, Russia\\
email:volovich@mi.ras.ru}

\date{}

\maketitle

\begin{abstract}
The eigenvalue problem for the square integrable solutions is
studied usually for elliptic equations. In this note we consider
such a problem for the hyperbolic Klein-Gordon equation on Lorentzian
manifolds. The investigation could help to answer the question why
elementary particles have a discrete mass spectrum.
 An infinite family of square integrable solutions for
the Klein-Gordon equation on the Friedman type manifolds is
constructed. These solutions have a discrete mass spectrum and a
finite action. In particular the solutions on   de Sitter space
are investigated.
\end{abstract}

\thispagestyle{empty}


\section{Introduction}

Let  $M$ be an $(n+1)$-dimensional manifold with a Lorentz metric
$g_{\mu\nu},~\mu,\nu=0,1,...,n$. Consider the Klein-Gordon
equation \cite{BD} on $M$ for the real valued function
$f$
\begin{equation}
\label{1} \square f+\lambda f=0.
\end{equation}
Here
$$
\square f=\nabla_{\mu}\nabla^{\mu}f=
\frac{1}{\sqrt{|g|}}\partial_{\mu}(\sqrt{|g|}g^{\mu\nu}\partial_{\nu}f),
$$
$g$ is the determinant of $(g_{\mu\nu})$ and the real parameter
$\lambda$ corresponds to the mass square.

We are interested in deriving the values of $\lambda$ for which
there exist classical solutions $f\in C^2(M),$ satisfying the
condition
\begin{equation}
\label{2}
 \int_{M}f^2\sqrt{|g|}dx<\infty
\end{equation}
The condition (\ref{2}) was first considered in \cite{Koz} for
solutions of the Klein-Gordon equation on de Sitter space. Let
us note that the condition (\ref{2}) includes the integration not
only over the spatial variables as it is done usually for the
quantum Klein-Gordon field \cite{BD} but also over the time-like
variable.

  To answer the question  why the consideration of the
requirement (\ref{2}) could be
interesting we present the following
motivations.

Understanding the mass spectrum of elementary particles is an outstanding
problem for  physics. Why the elementary particles  have their
observed pattern of masses? There is no answer to even a simpler question why
the mass spectrum is discrete?
We will show that  solutions of Eq. (\ref{1}) satisfy the condition
(\ref{2}) only for some discrete values of the parameter
$\lambda,$ i.e. we obtain quantization of masses. This is
interesting because the  mass in field theory is considered as an
arbitrary parameter but  in nature there is only a discrete set
of masses of elementary particles.
A finite mass spectrum was first obtained in \cite{Koz}
for  de Sitter space. The idea that the boundedness  of
the mass spectrum might be related with  de Sitter geometry in
the  momentum space is considered in \cite{Kad}.

We  show that there exist solutions which
satisfy the condition (\ref{2}) and moreover they have a {\it finite
action}. The requirement of the finiteness of the action in the
Lorentz signature is natural for example in the case when we study
the wave function of the Universe in real time in the
semiclassical approximation \cite{Vil}. It is  known that there are
solutions of some nonlinear equations with finite action
(instantons) but they exist only in the Euclidean time.
A symmetry which exploits
the feature that de Sitter and Anti de Sitter space are related by analytic
continuation is considered in \cite{HN}.

 Some mathematical
motivations are discussed in the conclusion.

The paper is composed as follows. In the next section square integrable
solutions of the Friedman type manifolds are constructed. Then
solutions on de Sitter space and on the Friedman space are considered.

\section{ Solutions on  the Friedman type manifolds}

Let us consider a manifold $M=I\times N^n$ with a metric:
\begin{equation}
\label{1A}
 ds^2=g_{\mu\nu}dx^{\mu}dx^{\nu}=dt^{2} -a^2(t)dl^2.
\end{equation}
Here  $I$ is an interval on the real axis, $I\subset \mathbb{R}$,
$~$ $a(t)$ is a smooth positive function on $I$, $~$ $N^n$ a
Riemannian manifold and
\begin{equation}
\label{1B} dl^2=h_{ij}(y)dy^i dy^j,~~i,j=1,...,n
\end{equation}
is a Riemannian metric on $N^n.$ Such manifolds $(M,g_{\mu\nu})$
will be called the Friedman type manifolds.

Eq. (\ref{1}) for the metric (\ref{1A})takes the form
\begin{equation}
\label{3}
 \ddot{f}+\frac{n}{a}\dot{a}\dot{f}-\frac{1}{a^2}\Delta_h
f+\lambda f=0
\end{equation}
where $\Delta_h$  is the Laplace-Beltrami operator fot the
metric $h_{ij}$,
\begin{equation}
\label{4} \Delta_h
f=\frac{1}{\sqrt{h}}\partial_{i}(\sqrt{h}h^{ij}\partial_{j}f)
 \end{equation}
and the condition (\ref{2}) reads
\begin{equation}
\label{2c}
 \int_{M}f^2\sqrt{|g|}dx=\int_{I\times
 N^n}f^2|a|^n\sqrt{h}dtdy<\infty
\end{equation}

Let $q\geq 0$ be the eigenvalue of the operator $-\Delta_h$ on
$N^n$ and $\Phi=\Phi(y)$ is the corresponding eigenfunction:
\begin{equation}
\label{5} -\Delta_h\Phi=q\Phi,
 \end{equation}

\begin{equation}
\label{6} \int_{N^n}\Phi^2\sqrt{h}dy<\infty
 \end{equation}
We set
\begin{equation}
\label{7} f=B(t)a(t)^{-\frac{n}{2}}\Phi(y).
 \end{equation}
 Then from (\ref{3}),(\ref{5}) we obtain the Sturm-Liouville (Schrodinger)
 equation
\begin{equation}
\label{9w} \ddot{B}+[\lambda - v(t) ]B=0
 \end{equation}
 where
\begin{equation}
\label{6AB}
 v(t)=\frac{n}{2}\frac{\ddot{a}}{a}+\frac{n}{2}(\frac{n}{2}-1)
 \frac{\dot{a}^2}{a^2}-\frac{q}{a^2}
 \end{equation}
We look for solutions $B(t)$ of Eq. (\ref{9w}) in $ L^2(I)$ since
for functions of the form (\ref{7}) the condition (\ref{2c}) takes
the form

$$
\int_{I}B^2 dt<\infty.
$$
Consider the case  $I= \mathbb{R}.$

{\bf Theorem 1.} {\it Let $M=\mathbb{R}\times N^n$ be the Friedman
type manifold with the metric of the form (\ref{1A}),(\ref{1B})
such that there exists a solution $\Phi$ of Eq. (\ref{5}) on $N^n$
which is not identically vanishing with an eigenvalue $q\geq 0$.
Let the smooth positive function  $a(t)$ on $\mathbb{R}$ is such
that $v(t)$ (\ref{6AB}) satisfies the condition
\begin{equation}
\label{6ABC}
 v(t)\rightarrow\infty ~~if ~~|t|\rightarrow\infty.
 \end{equation}

  Then for given $q,\Phi$
 (\ref{5}),(\ref{6}), the problem (\ref{1}),(\ref{2}) has an infinite family of
 solutions $f_j=B_j(t)a(t)^{-\frac{n}{2}}\Phi(y)$ with eigenvalues
 $\lambda_j,~j=1,2,...$  and moreover
 $\lambda_j\rightarrow\infty ~ when ~ j\rightarrow\infty.$ }

{\bf Proof.} It is a well known result by Weyl and Titchmarsh (see
for example \cite{Tit}, Sect.5.12-5.13) that under condition
(\ref{6ABC}) the Sturm-Liouville problem (\ref{9w}) has a
discrete spectrum in $ L^2(\mathbb{R})$. Then Theorem 1 follows.

 {\bf Example 1.} Let us take
\begin{equation}
\label{10} a(t)=C\exp(\alpha t^{2k}),~C>0,~\alpha >0,~k>1.
 \end{equation}
 Then
 $$
 v(t)=\frac{n}{2}[\alpha 2k(2k-1)t^{2k-2}+
 \frac{n}{2}(\alpha 2k)^2t^{4k-2}]-qC^{-2}\exp(-\alpha t^{2k})
 $$
In this case the condition (\ref{6ABC}) is satisfied. One has a
discrete spectrum $\lambda_1,\lambda_2,...$ and
 $\lambda_j\rightarrow\infty $~ when~ $j\rightarrow\infty.$

\section{Solutions on de Sitter space}

 For de Sitter space one has:
 $M=\mathbb{R}\times S^3$,
\begin{equation}
\label{9рp} ds^2=dt^{2} - \cosh^{2} t \cdot h_{ij}(y)dy^i dy^j,
 \end{equation}
 where $h_{ij}$ is the standard metric on the 3-dimensional sphere
  $S^3$. The eigenvalues of the operator  $-\Delta_h$ on the 3-sphere
  are equal to
 $q=j(j+2),~j=0,1,2,...$ and
\begin{equation}
\label{11a} v(t)=\frac{9}{4}-[\frac{3}{4}+j(j+2)]\frac{1}{\cosh^2
t}
 \end{equation}
 We set
\begin{equation}
\label{11G}
 \alpha=\frac{3}{4}+j(j+2), ~\nu^2=\frac{9}{4}-\lambda
 \end{equation}
Then Eq. (\ref{9w}) takes the form
\begin{equation}
\label{9r} \ddot{B}+[\frac{\alpha}{\cosh^2 t} -\nu^2 ]B=0
 \end{equation}

Theory of Eq. (\ref{9r}) is well known   \cite{Tit,Flu} and it was
used in \cite{Koz}
 to construct square integrable solution of the
 Klein-Gordon equation on de sitter space. Spectrum for positive values
 of  $\nu^2$ is discrete and for negative is continuous. We
 consider the first case, $\nu^2>0$.

 Eq.
(\ref{9r}) for $\alpha >0$ has a solution in $L^2(\mathbb{R})$ iff
\begin{equation}
\label{9р} 0<\nu=\frac{1}{2}(\sqrt{1+4\alpha}-1)-n, ~ n=0,1,2,...
 \end{equation}
 In our case, due to (\ref{11G}),
 $$
 0<\nu=j+\frac{1}{2}-n, ~~ j,n=0,1,2,...
 $$
There is a family of square integrable solutions of
 Eq. (\ref{9w}) with eigenvalues $\lambda$ of the form
\begin{equation}
\label{13q}
\lambda_{jn}=\frac{9}{4}-(j+\frac{1}{2}-n)^2,~~(j,n=0,1,2,...,j+\frac{1}{2}-n>0)
 \end{equation}
If $\lambda_{jn} \geq 0$ then we should have
 иметь
 $$
0<j+\frac{1}{2}-n\leq\frac{3}{2},~~ j,n=0,1,2,...
 $$
and therefore either $j=n$ and $\lambda_{jn}=2$, or $j=n+1$ and
$\lambda_{jn}=0$.

In the case $j=n,~ \nu=1/2$ for any $j=0,1,2,...$  Eq. (\ref{9r})
has a solution in  $L^2(\mathbb{R})$ of the form
 \begin{equation}
\label{14V} B_j(t)=\frac{1}{(\cosh t)^{1/2}}\sum_{s=0}^j \frac
{(-j)_s(j+2)_s}{(3/2)_s s!}\frac{1}{(e^{2t}+1)^s},
 \end{equation}
 $$
 (k)_0=1,~(k)_s=k(k+1)...(k+s-1).
 $$
In the case $j=n+1,~ \nu=3/2$  for any $j=1,2,...$ Eq. (\ref{9r})
has a solution in  $L^2(\mathbb{R})$  of the form
 \begin{equation}
\label{14VV} B_j(t)=\frac{1}{(\cosh t)^{3/2}}\sum_{s=0}^{j-1}
\frac {(1-j)_s(j+3)_s}{(5/2)_s s!}\frac{1}{(e^{2t}+1)^s}
 \end{equation}

 Let us denote  $H_{\lambda}$ the subspace in $L^2(M)$ formed by the
 square integrable solutions of Eq. (\ref{1}). We have proved the
 following

 \textbf{Theorem 2} (see \cite{Koz}).
 {\it Let $M=\mathbb{R}\times S^3$ be de Sitter space with the metric
 $$
 ds^2=dt^2-\cosh^2 t \cdot h_{ij}(y)dy^i dy^j.
 $$
 Then
 $\dim H_{\lambda}=\infty$  for all $\lambda = \lambda_{jn}$
 (\ref{13q}).
 Moreover, if $\lambda =\lambda_{jn}\geq 0$
 then either $\lambda =0,$ or $\lambda =2$.}

\section{Solutions on the Friedman space}

 {\bf 1.} In the inflation cosmology the following form of the Friedman-de Sitter
 metric is often used:
\begin{equation}
\label{13} ds^2=dt^2-e^{2Ht} \cdot h_{ij}(y)dy^i dy^j,
 \end{equation}
где $h_{ij}$ is a Riemannian metric on a compact 3-dimensional
manifold,
 $0<t<\infty$ and $0<H$ is  Hubble`s constant. In this case the
 function $v(t)$ (\ref{6AB}) is
 \begin{equation}
\label{14} v(t)=\frac{9}{4}H^2-qe^{-2Ht}
 \end{equation}
Eq. (\ref{9w}) on the semi-axis with boundary conditions
$B(0)=B(\infty)=0$ has an eigenvalue  in this case. If the
parameter $t$ is interpreted as the radius in spherical
coordinates then we get the known model of deuteron (see, for
example \cite{Flu}). The solution has the form
$$
B(t)=J_{\nu}(ce^{-Ht}),
$$
where
$$
c=\frac{\sqrt{q}}{H}>0,~~\nu=\frac{\sqrt{9H^2-4\lambda}}{2H}>0,
$$
and $J_{\nu}$ is the Bessel function. The eigenvalue $\lambda$ is
derived from the relation $J_{\nu}(c)=0.$

 {\bf 2.} The Friedman metric has the form
\begin{equation}
\label{13N} ds^2=dt^2-a^2(t) h_{ij}(y)dy^i dy^j
 \end{equation}
where $h_{ij}$ is a Riemannian metric on the manifold of positive,
negative or flat curvature.  The function $a(t)$ is derived from
the Einstein-Friedman equations
\begin{equation}
\label{13L} 3\dot{a}^2/a^2=8\pi \rho-3k/a^2,~~3\ddot{a}/a=
-4\pi(\rho+3p),
 \end{equation}
where $k=1, -1, 0$ for the manifolds for manifolds of the
positive, negative of flat curvature respectively. The pressure
$p$ and the mass density $\rho$ are related by an equation of
state $p=p(\rho)$. In particular, for massless thermal radiation
$(p=\rho/3)$ in a 3-dimensional torus $(k=0)$ one has
\begin{equation}
\label{13T} a(t)=c\sqrt{t},~~c>0,~~ 0<t<\infty.
 \end{equation}
In this case
\begin{equation}
\label{13TA} v(t)=-\frac{3}{16t^2}-\frac{q}{c^2t},~~q>0
 \end{equation}
and  the Sturm-Liouville equation (\ref{9w}) has a discrete
spectrum for negative $\lambda$ :
\begin{equation}
\label{13TB} \lambda_n=-\frac{4q^2}{c^4(4n+1)^2},~~n=1,2,...
 \end{equation}
Indeed, if we denote $\lambda =-\nu^2,~\nu >0$ and define a new
function $\varphi (x)$ by
$$
B(t)=e^{-\nu t}t^{1/4}\varphi(2\nu t)
$$
then from the Sturm-Liouville equation
$$
\ddot{B}(t)+[\lambda +\frac{3}{16t^2}+\frac{q}{c^2t} ]B(t)=0
$$
we obtain that the function $\varphi (x)$ satisfies the equation
for the degenerate hypergeometric function
$$
x\varphi^{\prime\prime}(x)+(\frac{1}{2}-x)\varphi^{\prime}(x)
-(\frac{1}{4}-\frac{q}{2\nu c^2})\varphi(x)=0.
$$
It is known that the last equation has  solutions with the required behavior at infinity
 only if
$$
\frac{1}{4}-\frac{q}{2\nu c^2}=-n,~~n=1,2,...
$$
which leads to (\ref{13TA}).

 For a compact 3-dimensional manifold of negative curvature
$(k=-1)$ the function $a(t)$ has the form
\begin{equation} \label{13LL} a(t)=\sqrt{t^2-c^2},~~0<c<t<\infty
 \end{equation}
and the corresponding  Sturm-Liouville problem also has a
discrete spectrum for negative $\lambda$ .

\section{Discussion and Conclusions  }

{\bf 1}. An action for the Klein-Gordon equation (\ref{1}) has
the form
 \begin{equation}
\label{15} S=\frac{1}{2}\int_M[(\nabla f,\nabla f)-\lambda
f^2]\sqrt{|g|}dx
 \end{equation}
where
$$
(\nabla f,\nabla f)=g^{\mu\nu}\partial_{\mu}f\partial_{\nu}f.
$$
On solutions of the form  (\ref{7}) on the Friedman type manifolds
the action takes the form
 \begin{equation}
\label{15B}  S=\frac{1}{2}\int_{I\times
 N^n}[(\dot{B}-\frac{n}{2}
 \frac{\dot{a}}{a}B)^2\Phi^2-a^{-2}B^2h^{ij}
 \partial_i\Phi\partial_j\Phi -\lambda B^2\Phi^2]dt\sqrt{h}dy
 \end{equation}
On the solutions on de Sitter space of the form (\ref{14V}),
(\ref{14VV}) the integral (\ref{15B}) is convergent, i.e. the
actio  is finite.

 {\bf 2.} Let us make the substitution into Eq.  (\ref{3})
$$
f=u(y,t)a(t)^{-\frac{n}{2}}
$$
Then we obtain the following equation
 \begin{equation}
\label{16} \ddot{u}-a(t)^{-2}\Delta_h u +[\lambda - w(t)]u=0
 \end{equation}
where
$$
w(t)=\frac{n}{2}\frac{\ddot{a}}{a}+\frac{n}{2}(\frac{n}{2}-1)
 \frac{\dot{a}^2}{a^2}
$$
We look for solutions satisfying the condition
$$
\int_{\mathbb{R}\times
 N^n}u(y,t)^2dt\sqrt{h}dy<\infty
$$
There exists a well developed spectral theory for elliptic
differential operators (see, for example \cite{Tit, DS}.
 There is a spectral theory of the Liouville operator in ergodic theory
 of dynamical systems \cite{AKN }. It would
be interesting to develop a spectral theory for hyperbolic
equations.

 Consider for example on the Schwartz space $S(\mathbb{R}^2)$
 of the functions $u=u(x,t)$ on the plane the hyperbolic
 differential operator of the form
\begin{equation}
\label{17} Au=\frac{\partial^2}{\partial
t^2}u-\frac{\partial^2}{\partial x^2} u +\phi(x,t)u
 \end{equation}
where the smooth real valued function $\phi(x,t)$ admits a power
bound  on the variables $x,t$ at infinity. The operator $A$ admits
a self-adjoint extension in $L^2(\mathbb{R}^2)$. In the particular
simple case when the function $\phi$ has the form $\phi=x^2-t^2,$
the operator   is the difference of the Schrodinger operators for
two harmonic oscillators. Hence it has in $L^2(\mathbb{R}^2)$ a
complete system of eigenfunctions
\begin{equation}
\label{17R} u_{jn}=H_j(t)H_n(x)\exp\{-\frac{1}{2}(t^2+x^2)\},
 \end{equation}

$$
Au_{jn}=
 \lambda_{jn}u_{jn},~~j,n=0,1,2,...,
$$
with the corresponding eigenvalues $\lambda_{jn}=2(n-j).$
Here$H_j$ are the Hermite polynomials.

The action
$$
S=\frac{1}{2}\int_{\mathbb{R}^2}(\dot{u}^2-u_x^2-\phi u^2+\lambda
u^2)dtdx
$$
is finite on solutions (\ref{17R}).

 {\bf 3.}  Note that together with  (\ref{1}) the so called
 equation with conformal coupling is considered:
\begin{equation}
\label{18} \square f+\xi Rf+\lambda f=0.
 \end{equation}
Here $R$ is the scalar curvature of the manifold $M$ and $ \xi
=(n-1)/4n.$  For de Sitter space $R=12,~ \xi=1/6$. In this case
Eq.  (\ref{18}) $\square f+(2+\lambda )f=0$ has square integrable
solutions if $2+\lambda =\lambda_{jn}$ (\ref{13q}) and in
particular if  $\lambda=0$.

In this note the square integrable solutions of the Klein-Gordon
equation for scalar field on manifolds have been considered. It
would be interesting to study square integrable solutions on more
general manifolds and also equations for fields with higher spins.


\section*{Acknowledgements}

V.K. is supported in part by RFBR grant  05-01-01058 and I.V. by RFBR grant
05-01-00884.


{\small

\begin{thebibliography}{99}

\bibitem{BD} \textit{N.D. Birrell and P.C.W. Davies.}
Quantum Fields in Curved Space, \\
Cambridge University Press, 1982.

\bibitem{Koz} \textit{V.V. Kozlov}. Square Integrable Solutions of the Klein - Gordon Equation
on\\
 de Sitter Space, Russian Mathematical Surveys, 1987, v. 42, N. 4, p. 171.

\bibitem{Kad} {\it V.G. Kadyshevsky}. Nucl. Phys. 1978, B141, p. 477;

{\it V.G. Kadyshevsky, M.D. Mateev, V.N. Rodionov, A.S. Sorin.}
hep-ph/0512332.

\bibitem{Vil} {\it A. Vilenkin}. Phys. Rev. 1989, D39, 1116.
\bibitem{HN} {\it G. 't Hooft, S. Nobbenhuis}. gr-qc/0602076.
\bibitem{Tit} \textit{E.C. Titchmarsh}. Eigenfunction Expansions Associated with
Second - Order Differential \\
Equations, Oxford, At the Clarendon
Press, 1946.

\bibitem{DS} \textit{N. Dunford and J.T. Schwartz.} Linear Operators Part II: Spectral Theory,
Self Adjoint \\
Operators in Hilbert Space. John Wiley \& Sons, New York, 1963.



\bibitem{Flu} \textit{S. Flugge.} Practical Quantum Mechanics, Springer, 1971.

\bibitem{AKN } \textit{V.I. Arnold, V.V. Kozlov, and
A.I. Neishtadt}. Mathematical Aspects of Classical and\\
 Celestial Mechanics. Springer-Verlag. 1993.

\end{thebibliography}
}
\end{document}